\documentclass{article}[11pt]
\usepackage{amssymb}
\usepackage{amsmath,amsthm}
\usepackage{tikz} 
\usetikzlibrary{arrows, automata, positioning,calc}
\usetikzlibrary{shapes,snakes}
\usepackage{color}

\DeclareSymbolFont{extraup}{U}{zavm}{m}{n}
\DeclareMathSymbol{\varheart}{\mathalpha}{extraup}{86} %
\DeclareMathSymbol{\vardiamond}{\mathalpha}{extraup}{87} %

\newcommand\mN{\mathbb N}

\newcommand\eg{\emph{e.g.}}
\newcommand\cf{\emph{cf.}}
\newcommand\ignore[1]{}

\newcommand{\domain}{\mathbb{D}}
\newcommand{\data}{\textup{data}}

\newcommand{\kqhide}[1]{\textcolor{teal}{sth old is hidden here}}

\newcommand{\A}{\mathcal{A}} 
\newcommand{\B}{\mathcal{B}} 
 
\newcommand{\locs}{\mathcal{L}}
\newcommand{\loc}{\ell}
\newcommand{\edges}{E}
\newcommand{\true}{\texttt{true}}
\newcommand{\init}{\textup{in}}
\newcommand{\acc}{\textup{acc}}

\newcommand{\config}{C}
\newcommand{\sconfig}{S} %

\newcommand{\sNodes}{\mathbb{S}} %
\newcommand{\sTo}{\Rightarrow} %

\renewcommand{\succ}{\textup{Succ}} 
\newcommand{\vect}[1]{\boldsymbol{#1}}

\newcommand{\indiscernible}[3]{{#1} \equiv_{#3} {#2}}

\newcommand{\register}{r}
\newcommand{\registers}{R}
\newcommand{\supp}{\textup{supp}}

\newcommand{\GURA}{\textup{GURA}}
\newcommand{\GRA}{\textup{GRA}}
\newcommand{\GNRA}{\textup{GNRA}}

\renewcommand{\frac}{\textup{frac}}

\newcommand\EXPSPACE{\ensuremath{\textsf{EXPSPACE}}}

\newtheorem{theorem}{Theorem}
\newtheorem{proposition}[theorem]{Proposition}
\newtheorem{fact}[theorem]{Fact}
\newtheorem{corollary}[theorem]{Corollary}
\newtheorem{example}{Example}

\advance \topmargin by -\headheight
\advance \topmargin by -\headsep
\evensidemargin \oddsidemargin
\begin{document}
\title{On the Containment Problem for Unambiguous Single-Register Automata with Guessing} 

\author{Antoine Mottet\\
Charles University Prague
\and Karin Quaas \\
Universit\"at Leipzig
}

\date{}
\maketitle
\begin{abstract}
Register automata extend classical finite automata with a finite set of registers that can store data from an infinite data domain for later equality comparisons with data from an input data word. 
While the registers in the original model of register automata, introduced in 1994 by Kaminski and Francez, can only store data occurring in the data word processed so far, 
we study here the more expressive class of register automata \emph{with guessing}, where registers can nondeterministically take any value from the infinite data domain, even if this data does not occur in the input data word. 
It is well known that the containment problem, i.e., the problem of deciding for two given register automata with guessing $\A$ and $\B$, whether the language $L(\A)$ accepted by $\A$ is contained in the language $L(\B)$ accepted by $\B$, is undecidable, even if $\B$ only uses a single register. 
We prove that the problem is decidable if $\B$ is unambiguous and uses a single register. 
\end{abstract}

\section{Introduction}
Register automata~\cite{DBLP:journals/tcs/KaminskiF94, KaminskiZeitlin} 
are a widely studied computational model that extend classical finite automata with finitely many \emph{registers} that can take values from an infinite set  and perform equality comparisons with data from the input word.  
Register automata accept \emph{data languages}, that is sets of \emph{data words} over $\Sigma\times\domain$, where $\Sigma$ is a finite alphabet, and $\domain$ is an infinite set called the \emph{data domain}. %

\begin{figure}[t]
	\centering
  \scalebox{.8}{ 
\begin{tikzpicture}[->,>=stealth',shorten >=1pt,auto,node distance=4cm,thick,node/.style={circle,draw,scale=0.9}, roundnode/.style={circle, draw=black, thick, minimum size=6mm},]
\tikzset{every state/.style={minimum size=0pt}};
\node[roundnode,initial, initial text={}]  (1) at (0,0) {$\loc_0$};
\node[roundnode]  (2) at (2,0) {$\loc_1$};
\node[roundnode,accepting]  (3) at (4,0) {$\loc_2$};
\path [->] (1) edge node[above] {\scriptsize{$\dot\register\neq$}} (2);
\path [->] (2) edge node[above] {\scriptsize{$=\register$}} (3);

\path [->] (2) edge [loop above] node[above] {\scriptsize{$\neq\register, \dot\register=\register$}} (2);

\ignore{
\node  (c) at (10,2) {$\{(\loc_0,\bot)\}$};
\node  (c1) at (8.5,1) {$\{(\loc_1,d) \mid d\in\domain_\bot\backslash\{1\}\}$};

\node  (c11) at (8,0) {$\{(\loc_1,d) \mid d\in\domain_\bot\backslash\{1\}\}$};
\node  (c12) at (13,0) {$\{(\loc_1,d) \mid d\in\domain_\bot\backslash\{1,2\}\}\cup\{(\loc_2,2)\}$};

\node  (c111) at (7.5,-1) {$\{(\loc_1,d) \mid d\in\domain_\bot\backslash\{1\}\}$};
\node  (c112) at (9.8,-1) {$\dots$};
\node  (c123) at (13,-1) {$\{(\loc_1,d) \mid d\in\domain_\bot\backslash\{1,2,3\}\}\cup\{(\loc_2,3)\}$};
\node  (c132) at (16.4,-1) {$\dots$};
\node  (c1234) at (14.5,-1.7) {$\dots$};
\node  (c1234) at (7.5,-1.7) {$\dots$};

\node  (c3) at (11.8,1.6) {$\dots$};
\path [->] (c) edge node[right] {\scriptsize{$\ 1$}} (c1);
\path [->] (c1) edge node[right] {\scriptsize{$\ 1$}} (c11);
\path [->] (c11) edge node[right] {\scriptsize{$\ 1$}} (c111);
\path [->] (c12) edge node[right] {\scriptsize{$\ 3$}} (c123);
}
 	\end{tikzpicture}  
  }
\caption{
	A $\GURA$ with a single register $\register$ and over a singleton alphabet (we omit the labels at the edges). %
}
\label{fig:gura}
\end{figure}
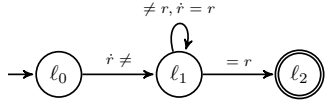

As an example, consider the register automaton in Figure \ref{fig:gura} using a single register $\register$ ($\dot r$ refers to the future value of $r$). 
This automaton processes finite data words over $\Sigma\times\domain$. 
We assume that $\Sigma=\{\sigma\}$ is a singleton, so that we omit the letter $\sigma$ from all transitions and input words, and $\domain=\mN$. %
Let us study the behaviour of the automaton: 
starting in  the initial location $\loc_0$ and processing the first input letter $d$, 
the automaton can only move to $\loc'$ if it satisfies the register constraint $\dot\register \neq$. 
This constraint requires the register $\register$, when reaching $\loc_1$, to store a data value $d'\in\mN$ such that $d'\neq d$. 
The automaton can nondeterministically \emph{guess} such a datum $d'$. 
Being in $\loc_1$ with the register holding the value $d'$, by the constraint $=\register$, it can only move to the accepting location $\loc_2$ if it reads the input letter $d'$; 
for every other input letter, satisfying the constraint $\neq \register$, 
the automaton stays in $\loc_1$, keeping the register value (indicated by the constraint $\dot\register=\register$). 
For instance, 
for the input data word $w_\mathit{ad}=1 \, 2 \, 2 \, 3$, 
there are infinitely many distinct runs (one for each guessed datum different from $1$), 
but only one accepting run, namely
$$(\loc_0,\bot)\xrightarrow{1}(\loc_1,3)\xrightarrow{2}(\loc_1,3)\xrightarrow{2}(\loc_1,3)\xrightarrow{3}(\loc_2,3).$$
We write  
$L_{\mathit{ad}} = \{d_1\dots d_k\mid \forall  k\geq 2 \, \,  \forall 1\leq i<k. \, \,  d_i\neq d_k\}$
to denote the set of data words that is accepted by the automaton in Figure \ref{fig:gura} ($\mathit{ad}$ standing for \emph{all different}).

We remark that the nondeterministic \emph{guessing} of data values to store them into registers for future comparisons is not allowed in the original model of register automata, introduced by Kaminski and Francez~\cite{DBLP:journals/tcs/KaminskiF94}, and studied \eg \ in~\cite{DBLP:journals/tocl/DemriL09,DBLP:conf/stacs/MottetQ19,DBLP:journals/tocl/QuaasS19}. 
In fact, the model that we study here is strictly more expressive (with respect to acceptance of data languages) than the classical model. 
In order to distinguish the model with guessing from the classical model without guessing, we explicitly refer to the former by \emph{register automata with guessing}, \GRA \ for short.

The \GRA \  in Figure \ref{fig:gura} is not \emph{deterministic}: 
being in $\loc_0$ and processing the first input datum $d$, it can nondeterministically \emph{guess} any datum $d'$ such that $d'\neq d$ for storage in $\register$. 
However, one can easily see that for every input data word there is \emph{at most one accepting run}, uniquely determined by the single guessed datum  $d'$. 
We call automata that have, for every input word, at most one accepting run, \emph{unambiguous}. 
One of the main open problems concerning unambiguous register automata with guessing (GURA, for short) is whether the class of data languages accepted by GURA is closed under complementation\footnote{In Theorem 12 in~\cite{DBLP:conf/dcfs/Colcombet15}, it is claimed that the class of data languages accepted by GURA is effectively closed under complement; however, to the best of our knowledge, this claim remains unproved.}. 
(In contrast, it is known that data languages accepted by unambiguous register automata (without guessing) are \emph{not} closed under complementation; for instance, the complement of $L_{\mathit{ad}}$  can be accepted by an unambiguous register automata with a single register, but $L_{\mathit{ad}}$ cannot even be accepted by any nondeterministic register automaton (without guessing)~\cite{KaminskiZeitlin}.)

In this paper, we study the \emph{containment problem}: given two $\GRA$ $\A$ and $\B$, does $L(\A)\subseteq L(\B)$ hold? Here, $L(\A)$ and $L(\B)$, respectively, denote the set of data words accepted by $\A$ and $\B$, respectively. This problem, playing a central role in formal verification, has been studied a lot for register automata, see \eg~\cite{DBLP:journals/tocl/NevenSV04,DBLP:journals/tocl/DemriL09,DBLP:conf/stacs/MottetQ19}. 
For $\GRA$, it is well known that the problem is undecidable~\cite{DBLP:journals/tocl/DemriL09}. More detailed, the special case of deciding whether a single given $\GRA$ $\B$ over $\Sigma$ and $\domain$ accepts the set $(\Sigma\times\domain)^*$ of all data words, is undecidable, even if $\B$ only uses a single register\footnote{A proof for undecidability can be done using a reduction from the undecidable reachability problem for Minsky machines, following the lines of the proof of Theorem 5.2 in~\cite{DBLP:journals/tocl/DemriL09}. The nondeterministic guessing can be used to express that there exists some decrement for which there is no matching preceding increment.}.

In this short note, we prove that the containment problem $L(\A)\subseteq L(\B)$ is decidable in $\EXPSPACE$ if $\B$ is unambiguous and uses a single register (and no restriction on $\A$).

\section{Main Definitions}
In this section, 
we define \emph{register automata with guessing} as introduced by Kaminski and Zeitlin~\cite{KaminskiZeitlin}. 

We start with some preliminary notions. 
We use $\Sigma$ to denote a finite alphabet, and $\domain$ to denote an infinite data domain. %
A \emph{data word} is a finite sequence $(\sigma_1,d_1)\dots(\sigma_k,d_k) \in (\Sigma\times\domain)^*$. We use $\varepsilon$ to denote the empty data word. 
A \emph{data language} is a set of data words. 
We use $\data(w)$ to denote the set $\{d_1,\dots,d_k\}$ of all data occurring in $w$.

Let $\domain_\bot$ denote the set $\domain\cup\{\bot\}$, where $\bot\not\in\domain$. We let $\bot \neq d$ for all $d\in\domain$. 
We use boldface lower-case letters like $\vect{a}, \vect{b}, \dots $ to denote tuples in $\domain_\bot^n$, where $n\in\mN$.
Given a tuple $\vect{a}\in\domain_\bot^n$, we write $a_i$ for its $i$-th component, and $\data(\vect{a})$ denotes the set $\{a_1,\dots,a_n\}\subseteq\domain_\bot$ of all data  occurring in $\vect{a}$.

Let $\registers=\{\register_1,\dots,\register_n\}$ be a finite set of \emph{registers}. 
A \emph{register valuation} is a mapping $\vect{u}:\registers\to\domain_\bot$; we may write $u_i$ as shorthand for $\vect{u}(\register_i)$. Let $\domain_\bot^\registers$ denote the set of all register valuations. 
A \emph{register constraint over $\registers$} is defined by the grammar
\begin{align*}
\phi ::= 
\true \,\mid\, 
\, =\register_i \,\mid\,
\dot\register_i=\register_j \,\mid\,  
\dot\register_i= \, \,\mid\, 
\neg\phi \,\mid\, 
  \phi \wedge \phi 
\end{align*} 
where $\register_i,\register_j\in\registers$. 
Intuitively, $\register_i$ refers to the current value of the register $\register_i$, while $\dot\register_i$ refers to the future value of the register $\register_i$.  
We use $\Phi(\registers)$ to denote the set of all register operations over $\registers$.  
The satisfaction relation $\models$ on $\domain_\bot^\registers\times\domain\times\domain_\bot^\registers$ is defined by structural induction as follows. We only give the atomic cases; the  cases for the Boolean formulas are as usual. We have  $(\vect{u},d,\vect{v}) \models \phi$ if 
\begin{itemize}
\item $\phi$ is of the form $\true$,
\item $\phi$ is of the form $=\register_i$  and $u_i = d$, 
\item $\phi$ is of the form $\dot\register_i=\register_j$  and $v_i = u_j$,  
\item $\phi$ is of the form $\dot\register_i=$ and $v_i = d$. 
\end{itemize}
For instance, $(1,2,1)\models (\neq\register)\wedge(\dot\register=\register)$, while $(1,2,3)$ does not. 
Note that only register constraints of the form $\dot\register_i=\register_j$ and $\dot\register=$ 
uniquely determine the new value of $\register_i$.
In absence of such a register constraint, the new value of $\register_i$ can be equal to  
(almost\footnote{The register constraint $\dot\register\neq$ requires that the new value of $\register$ is different from the current input datum, so that $\register$ may take any of the infinitely data in $\domain$ except for the input datum. Likewise, the register constraint $\dot\register_i\neq\register_j$ requires that $\register_i$ takes any of the infinitely data in $\domain$ except for the current value of $\register_j$.}) 
any of the infinitely many data values in $\domain$. 
Register automata that allow for such nondeterministic \emph{guessings} of future register values are called \emph{register automaton with guessing}. 
Formally, a register automaton with guessing (GRA) over $\Sigma$ is a tuple $\A = (\registers,\locs, \loc_{\init}, \locs_{\acc}, \edges)$, where
\begin{itemize}
\item $\registers$ is a finite set of registers, 
\item $\locs$ is a finite set of locations, 
\item $\loc_{\init}\in\locs$ is the initial location, 
\item $\locs_{\acc}\subseteq\locs$ is the set of accepting locations, 
\item $\edges\subseteq\locs\times\Sigma\times\Phi(\registers)\times\locs$ is a finite set of edges.
\end{itemize}
A \emph{state} of $\A$  is a pair $(\loc,\vect{u})\in \locs\times\domain_\bot^\registers$, where $\loc$ is the current location and $\vect{u}$ is the current register valuation.  
Given two states $(\loc,\vect{u})$ and $(\loc',\vect{u'})$ and some input letter $(\sigma,d)\in\Sigma\times\domain$, we postulate a transition $(\loc,\vect{u})\xrightarrow{\sigma,d}_\A(\loc',\vect{u'})$ if there exists some edge $(\loc,\sigma,\phi,\loc')\in\edges$ such that $(\vect{u},d,\vect{u'})\models\phi$. 
A \emph{run} of $\A$ on the data word 
$(\sigma_1,d_1)\dots(\sigma_k,d_k)$ is a  sequence $(\loc_0,\vect{u^0}) \xrightarrow{\sigma_1,d_1}_\A (\loc_1,\vect{u^1}) \xrightarrow{\sigma_2,d_2}_\A \dots \xrightarrow{\sigma_k,d_k}_\A (\loc_n,\vect{u^k})$ of such transitions. 
We say that a run like above \emph{starts in $(\loc,\vect{u})$} if $(\loc_0,\vect{u^0})=(\loc,\vect{u})$. 
A run is \emph{initialized} if starts in $(\loc_\init,\{\bot\}^k)$, 
and a run is \emph{accepting} if $\loc_k\in\locs_\acc$. 
The data language \emph{accepted} by $\A$, denoted by $L(\A)$, is the set of data words for which there exists an initialized, accepting run of $\A$. 
A $\GRA$ is unambiguous ($\GURA$) if for every input data word $w$ there is at most one initialized accepting run. 
The \emph{containment problem} is the following decision problem: given two $\GRA$ $\A$ and $\B$, does $L(\A)\subseteq L(\B)$ hold?

\section{Some Facts about Register Automata}
\subsection{Unambiguous Register Automata with Guessing}
Fix a GURA $\B=(\registers,\locs,\loc_{\init},\locs_\acc,\edges)$ with a single register $\register$. 
Let $\config\subseteq (\locs\times\domain_\bot)$ be a set of states of $\B$, and let $(\sigma,d)\in(\Sigma\times\domain)$. 
We use $\succ_\B(\config,(\sigma,d))$ to denote the \emph{successor of $\config$ on the input $(\sigma, d)$}, formally defined by 
\begin{align*}
\succ_\B(\config,(\sigma,d)) := \{(\loc,u) \in (\locs\times\domain_\bot) \mid \exists (\loc',u')\in \config. (\loc',u')\xrightarrow{\sigma,d}_\B(\loc,u)\}.
\end{align*}
In order to extend this definition to data words, we define inductively $\succ_\B(\config,\varepsilon):=\config$ and $\succ_\B(\config, w \cdot (\sigma,d))  :=  \succ_\B(\succ_\B(\config,w),(\sigma,d))$. 
We say that a set $\config\subseteq (\locs\times\domain)$ of states is \emph{reachable in $\B$} if there exists some data word $w$ such that $\config = \succ_\B(\config_\init, w)$, where $\config_\init=\{(\loc_\init,\bot)\}$.   

A \emph{configuration of $\B$} is a finite union of finite or cofinite subsets of $\locs\times\domain_\bot$. Hence the set $\config_\init:=\{(\loc_\init,\bot)\}$ is a configuration, henceforth called the \emph{initial configuration}. 
Note that for all configurations $\config$ and data words  $w$, the successor $\succ_\B(\config,w)$ is a configuration, too. 
This implies that every reachable set $\config\subseteq (\locs\times\domain)$ of states is a configuration.
Given a configuration $\config$, we use $\data(\config)$ to denote the set $\{d\in\domain_\bot \mid \exists \loc\in\locs. (\loc,d)\in\config\}$ of data occurring in $\config$. 

The \emph{support} of a configuration $\config$ is the set $\supp(C)$ of data $d$ such that at least one of the following holds:
\begin{itemize}
\item $(\loc, d)\in\config$ for some $\loc$ such that $(\{\loc\}\times \domain)\cap\config$ is finite,
\item $(\loc,d)\not\in\config$ for some $\loc$ such that $(\{\loc\}\times \domain)\cap\config$ is cofinite.
\end{itemize}

We say that a configuration $\config$ is \emph{coverable} if there exists some configuration $\config'\supseteq \config$ such that $\config'$ is reachable in $\B$. 
We say that a configuration $\config$ is \emph{accepting} if there exists $(\loc,u)\in\config$ such that $\loc\in\locs_\acc$; otherwise we say that $\config$ is \emph{non-accepting}. 
The following proposition follows immediately from the definition of \GURA.
\begin{proposition}
\label{prop:gura_implied_badness}
If $\config,\config'$ are two configurations of $\B$ such that $\config\cap\config'=\emptyset$ and $\config\cup\config'$ is coverable, then for every data word $w$ the following holds: 
if $\succ_\B(\config,w)$ is accepting, then $\succ_\B(\config',w)$ is non-accepting. 
\end{proposition}
A \emph{partial isomorphism of $\domain_\bot$} is an injective mapping $\pi:D\to\domain_\bot$ with
domain $\textup{dom}(\pi):=D\subseteq\domain$ such that if $\bot\in D$ then $\pi(\bot)=\bot$. 
Let $\pi$ be a partial isomorphism of $\domain_\bot$ and let $\config$ be a configuration such that $\data(\config)\subseteq\textup{dom}(\pi)$.
We define the configuration $\pi(\config) := \{(\loc,\pi(d))\mid (\loc,d)\in\config\}$; likewise, if $\{d_1,\dots,d_k\}\subseteq\textup{dom}(\pi)$, we define the data word  $\pi(w)=(\sigma_1,\pi(d_1))\dots(\pi_k,\pi(d_k))$. 
We say that $\config, w$ and $\config',w'$ \emph{are equivalent with respect to $\pi$}, written $\config,w \sim_\pi \config',w'$, if
$\pi(\config)=\config' \text{ and } \pi(w)=w'$. If $w=w'=\varepsilon$, we may write $\config\sim_\pi\config'$. We write $\config,w\sim\config',w'$ if $\config,w\sim_\pi\config',w'$ for some partial isomorphism $\pi$ of $\domain_\bot$. 
\begin{proposition}
\label{prop:ra_equivalence}
If $\config,w\sim\config',w'$, then
$\succ(\config,w) \sim \succ(\config',w')$. 
\end{proposition}
As an immediate consequence of Proposition \ref{prop:ra_equivalence}, we obtain that $\sim$ preserves the configuration properties of being \emph{accepting} respectively \emph{non-accepting}. 
\begin{corollary}
\label{corollary:gnra_bad}
If $\config,w\sim\config',w'$ and $\succ_\B(\config,w)$ is non-accepting (accepting, respectively), then $\succ_\B(\config',w')$ is non-accepting (accepting, respectively). 
\end{corollary}
Combining the last corollary with Proposition \ref{prop:gura_implied_badness}, we obtain 
\begin{corollary}
\label{corollary:gura_bad}
If $\config,\config'$ are two configurations such that $\config\cap\config'=\emptyset$ and $\config\cup\config'$ is coverable in $\B$, then 
for every data word $w$ such that $\config,w\sim\config',w$, the configurations 
$\succ_\B(\config,w)$ and $\succ_\B(\config',w)$ are non-accepting. 
\end{corollary}

\subsection{The Synchronized State Space}
For the rest of this paper, let  $\A=(\registers^\A,\locs^\A,\loc^\A_{\init},\locs^\A_{\acc},\edges^\A)$ be a \GNRA \ over $\Sigma$ with $\registers^A=\{\register_1,\dots,\register_m\}$, and let 
$\B=(\registers^\B,\locs^\B,\loc^\B_{\init},\locs^\B_{\acc},\edges^\B)$ be a \GURA \ over $\Sigma$ with a single register $\register$. 
A \emph{synchronized configuration of $\A$ and $\B$} is a pair 
$((\loc,\vect{d}),\config)$, where $(\loc,\vect{d})\in (\locs^\A\times\domain^{\registers^\A}_\bot)$ is a single state of $\A$, and
$\config\subseteq (\locs^\B\times\domain_\bot)$ is a configuration of $\B$.  
We define $\sconfig_{\init}:=((\loc_{\init}^\A,\{\bot\}^m), \config_\init)$ to be the \emph{initial synchronized configuration of $\A$ and $\B$}.
We define the \emph{synchronized state space of $\A$ and $\B$} to be the  (infinite) state-transition system $(\sNodes,\sTo)$, where $\sNodes$ is the set of all synchronized configurations of $\A$ and $\B$, and $\sTo$ is defined as follows. If $\sconfig=((\loc,\vect{d}),\config)$ and $\sconfig'=((\loc',\vect{d'}),\config')$, then
$\sconfig\sTo\sconfig'$ if there exists a letter $(\sigma,d)\in(\Sigma\times\domain)$ such that $(\loc,\vect{d})\xrightarrow{\sigma,d}_\A(\loc',\vect{d'})$, and 
$\succ_\B(\config,(\sigma,d))=\config'$. 
We say that a synchronized configuration \emph{$\sconfig$ reaches a synchronized configuration $\sconfig'$ in $(\sNodes,\sTo)$} if there exists a path in $(\sNodes,\sTo)$ from $\sconfig$ to $\sconfig'$. 
We say that a synchronized configuration $\sconfig$ is \emph{reachable in $(\sNodes,\sTo)$} if $\sconfig_\init$ reaches $\sconfig$. 
We say that a synchronized configuration $\sconfig=((\loc,\vect{d}),\config)$ is \emph{coverable in $(\sNodes,\sTo)$} if there exists some configuration  $\config'\supseteq\config$  such that $((\loc,\vect{d}),\config')$ is reachable in $(\sNodes,\sTo)$. 

We aim to reduce the containment problem $L(\A)\subseteq L(\B)$ to a reachability problem in $(\sNodes,\sTo)$. For this, call a  synchronized configuration $((\loc,\vect{d}),\config)$ \emph{bad} if $\loc\in\locs_\acc^\A$ is an accepting location and $\config$ is non-accepting, i.e., $\loc'\not\in\locs_\acc^\B$ for all $(\loc',u)\in\config$. 
The following proposition is easy to prove, cf.~\cite{DBLP:conf/lics/OuaknineW04}. 

\begin{proposition}
\label{prop:reductionToReach}
$L(\A)\subseteq L(\B)$ does not hold if, and only if, some bad synchronized configuration is reachable in $(\sNodes,\sTo)$. 
\end{proposition} 

We extend the equivalence relation $\sim$ defined above to synchronized configurations in a natural manner, i.e,  given a partial isomorphism $\pi$ of $\domain_\bot$
such that $\data(\vect{d})\cup\data(\config)\subseteq\textup{dom}(\pi)$,
we define $((\loc,\vect{d}),\config) \sim_\pi ((\loc,\vect{d}'),\config')$ if $\pi(\config)=\config'$ and $\pi(\vect{d})=\vect{d}'$.
We shortly write $\sconfig\sim\sconfig'$ if there exists a partial isomorphism $\pi$ of $\domain_\bot$ such that $\sconfig\sim_\pi\sconfig'$.
Clearly, an analogon of Proposition~\ref{prop:ra_equivalence} holds for this extended relation. In particular, we have the following:
\begin{proposition}\label{prop:equivalence-relation-synch-compatible}
Let $\sconfig,\sconfig'$ be two synchronized configurations of $(\sNodes,\sTo)$ such that $\sconfig\sim\sconfig'$.
If $\sconfig$ reaches a bad synchronized configuration, so does $\sconfig'$.
\end{proposition}
Note that $(\sNodes,\sTo)$ is infinite so that \emph{a priori} it is not clear how to exploit Proposition \ref{prop:reductionToReach} to solving the containment problem. First of all, $(\sNodes,\sTo)$ is not finitely branching: for every synchronized configuration $\sconfig=((\loc,\vect{d}),\config)$ in $\sNodes$, every input datum $d\in\domain$ and every \emph{guessed} new value of the registers may give rise to its own individual synchronized configuration $\sconfig_d$ such that $\sconfig\sTo\sconfig_d$. However, it is well known that, using standard techniques, one can define an \emph{abstract} finite-branching  state-transition system that is bisimilar to $(\sNodes,\sTo)$ with respect to $\sim$, \cf~\cite{DBLP:conf/stacs/MottetQ19}.  
Second, and potentially more harmful, the data needed to define the configuration $\config$ in a synchronized configuration $((\loc,\vect{d}),\config)$ can grow unboundedly. As an example, consider the $\GURA$ in Figure \ref{fig:gura}. For every $k\geq 1$, the configuration $\{(\loc_1,d)\mid d\in\mN\backslash\{d_1,\dots,d_k\}\}\cup\{(\loc_2,d_k)\}$ with pairwise distinct data values $d_1,\dots,d_k$ is reachable by inputting the data word $d_1 \, \dots d_k$. 
In the next section, we prove that one can solve the reachability problem from Proposition \ref{prop:reductionToReach} by focussing on a subset of configurations of $\B$ that can be defined by a bounded number of data. 
The approach follows the ideas presented in~\cite{DBLP:conf/stacs/MottetQ19} for unambiguous register automata (without guessing); 
however, the main technical proposition in~\cite{DBLP:conf/stacs/MottetQ19} does not apply to $\GRA$ and is substituted by Proposition \ref{prop:collapse_gura} below. 

\section{Decidability of the Containment Problem}

\subsection{Bounding the Size of the Supports}

Recall the equivalence relation $\sim$ on $k$-tuples, where for $\vect a,\vect b\in\domain_\bot^k$ we have $\vect a\sim \vect b$ if there exists a partial isomorphism $\pi$ of $\domain_\bot$
such that $\pi(\vect a)=\vect b$. 
Note that this equivalence relation has finitely many equivalence classes for all $k\in\mN$. 
As an example, for $k=3$ the equivalence classes of triples are the classes of $(d_0,d_0,d_0),(d_0,d_0,d_1),(d_0,d_1,d_0),(d_1,d_0,d_0),(d_0,d_1,d_2)$, where $d_0,d_1,d_2$ are pairwise distinct data values.

Let $\sconfig=((\loc,\vect{d}),\config)$ be a synchronized configuration, and let $a,b\in\supp(\config)$ be two data values in the support of $\config$.  
We say that \emph{$a$ and $b$ are indistinguishable in $\sconfig$}, written $\indiscernible{a}{b}{\sconfig}$, 
if $a,b\not\in\data(\vect{d})$ and $\{\loc \in\locs \mid (\loc,a)\in C\} = \{\loc\in\locs\mid (\loc,b)\in C\}$.

Let $\sconfig$ be a synchronized configuration $((\loc,\vect d),\config)$ and let $a,b\in\supp(\config)\setminus\data(\vect d)$.

Given a configuration $\config$, we define for every datum $d\in\domain$ the sets
\begin{align*}
\config_d^+ := & \, \{(\loc,d)\in \locs\times\{d\} \mid (\loc,d)\in\config \text{ and } \data(\config\cap (\{\loc\}\times\domain)) \text{ is finite}\}, \text{ and} \\
\config_d^- := & \, \{(\loc,d)\in \locs\times\{d\} \mid (\loc,d)\not\in\config \text{ and } \data(\config\cap (\{\loc\}\times\domain)) \text{ is infinite}\}
\end{align*}
For later reference, we state the following simple fact. 
\begin{fact}
\label{fact:minusandconfignointersect}
$\config\cap \config^-_d=\emptyset$, for all configurations $\config$ and data $d\in\domain$.
\end{fact}
\begin{example}
Let $\config = \{(\loc_1,0),(\loc_1,1)\} \cup \{(\loc_2,d) \mid d\in\mN\backslash\{1,2\}\}\cup\{(\loc_3,d)\mid d\in\mN\backslash\{0,1\}\}$. Then
\begin{center}
\begin{tabular}{lll}
$\config^+_0 = \{(\loc_1,0)\}$ & $\config^+_1 = \{(\loc_1,1)\}$ & $\config^+_2=\emptyset$ \\
$\config^-_0=\{(\loc_3,0)\}$ & $\config^-_1 = \{(\loc_2,1),(\loc_3,1)\}$ & $\config^-_2=\{(\loc_2,2)\}$
\end{tabular}
\end{center}
\end{example}
We say that a configuration $\config$ is \emph{essentially coverable} if for every two $(\loc,u),(\loc',u')\in\config$, the set $\{(\loc,u),(\loc',u')\}$ is coverable. 
\begin{proposition}
\label{prop:esscov}
Let $\config$ be an essentially coverable configuration, and let $b\in\supp(\config)$.
Then $((\config\backslash\config^+_b)\cup\config^-_b)$ is essentially coverable, too. \end{proposition}
\begin{proof}
Let $(\loc,c),(\loc',c')\in ((\config\backslash\config^+_b)\cup\config^-_b)$. 
If $(\loc,c),(\loc',c')\in\config\backslash\config^+_b$, then
$\{(\loc,c),(\loc',c')\}$ is coverable by essential coverability of $\config$. 
Suppose $(\loc,c),(\loc',c')\in\config^-_b$. By definition of $\config^-_b$, $c=c'=b$. 
Pick some value $e\in\domain\backslash\{b\}$ such that $(\loc,e),(\loc',e)\in\config$. Note that such a value $e$ must exist, as by definition of $\config^-_b$, the sets $\data((\{\loc\}\times\domain)\cap C)$ and $\data((\{\loc'\}\times\domain)\cap C)$ are cofinite, and hence their intersection is non-empty.
By essential coverability of $\config$, $\{(\loc,e),(\loc',e)\}$ is coverable. 
There must thus exist some data word $w$ such that $\{(\loc,e),(\loc',e)\}\subseteq \succ((\loc_\init,\bot),w)$. 
Let $\pi$ be any partial isomorphism satisfying $\pi(e)=b$ and whose domain contains $\data(w)$. 
Clearly, $\{(\loc,b),(\loc',b)\} \subseteq \succ((\loc_\init,\bot),\pi(w))$, and hence $\{(\loc,b\},(\loc',b)\}$ is coverable. 
Finally, suppose $(\loc,c)\in\config\setminus C_b^+$ and $(\loc',c')\in\config^-_b$. The proof that $\{(\loc,c),(\loc',c')\}$ is coverable is very similar to the proof for the preceding case and left as an exercise. 
\end{proof}

\begin{proposition}\label{prop:collapse_gura}
Let $\sconfig=((\loc^\A,\vect{d}),\config)$ be a synchronized configuration of $\A$ and $\B$ such that $\config$ is essentially coverable, and let $a\neq b$ be such that $a,b\in\supp(\config)$ and $\indiscernible{a}{b}{\sconfig}$.
	$\sconfig$ reaches a bad configuration in $(\sNodes,\sTo)$ if, and only if, $S':=((\loc^\A,\vect{d}),(\config\setminus \config_b^+)\cup \config_b^-)$ reaches a bad configuration in $(\sNodes,\sTo)$.
\end{proposition}
\begin{proof}
	$(\Leftarrow)$
	Suppose there exists some data word $w$ such that
	 there exists an accepting run of $\A$ on $w$ that starts in $(\loc^\A,\vect{d})$,  and $\succ_\B(\config\backslash \config_b^+\cup \config_b^-,w)$ is non-accepting. 
	We assume in the following that $\succ_\B(C^+_b,w)$ is accepting; otherwise we are done. 
	Let $(\loc^+,b)\in C_b^+$ be the unique state such that $\succ_\B((\loc^+,b),w)$ is accepting. 	
	In the following, we prove that we can without loss of generality assume that $w$ does not contain any $a$'s. 
	Pick some $a'\in\domain$ such that $a'\not\in\data(w)\cup\supp(\config)\cup\data(\vect{d})$. 
	Let $\pi$ be the isomorphism defined by $\pi(a)=a'$, $\pi(a')=a$, and $\pi(d)=d$ for all $d\in\domain_\bot\backslash\{a,a'\}$.  
	Then 
	$(\loc^\A,\vect{d}),w \sim_\pi(\loc^\A,\vect{d}),\pi(w)$ (as $a\not\in\data(\vect{d})$ by $\indiscernible{a}{b}{\sconfig}$), and 
	$(\loc^+,b),w \sim_\pi (\loc^+,b),\pi(w)$.  
	By Corollary \ref{corollary:gnra_bad}, there exists an accepting run of $\A$ on $\pi(w)$ that starts in $(\loc^\A,\vect{d})$, and  $\succ_\B((\loc^+,b),\pi(w))$ is accepting. 
	We prove that $\succ_\B((\loc,c),\pi(w))$ is non-accepting, for every $(\loc,c)\in\config\setminus\{(\loc^+,b)\}\cup\config^-_b$:  
	first, let $(\loc,c)\in C\setminus\{(\loc^+,b)\}$. 
	By essential coverability of $\config$, $\{(\loc^+,b),(\loc,c)\}$ is coverable. 
	By Proposition \ref{prop:gura_implied_badness}, 
	$\succ_\B((\loc,c),\pi(w))$ must be non-accepting. 
	Second, let $(\loc,c)\in\config^-_b$. But then $c=b$, and hence $(\loc,c),w \sim_\pi (\loc,c),\pi(w)$. By assumption, $\succ_\B((\loc,c),w)$ is non-accepting, so that by Corollary \ref{corollary:gnra_bad},  $\succ_\B((\loc,c),\pi(w))$ is non-accepting, too. 
	Note that $\pi(w)$ indeed does not contain any $a$'s. 
	We can hence continue the proof assuming that 
	$w$ does not contain any $a$'s. 
	
	Next, we prove that if we replace all $b$'s occurring in $w$ by some fresh datum not occurring in $\supp(\config)\cup\data(w)\cup\data(\vect{d})$, we 
	obtain a data word that guides $\sconfig$ to a bad synchronized configuration. 
	Formally, pick some datum $b'\not\in\data(w)\cup\supp(\config)\cup\data(\vect{d})$, and let $\pi$ be the isomorphism defined by 
	$\pi(b)=b'$, $\pi(b')=b$, and $\pi(d)=d$ for all $d\in\domain_\bot\backslash\{b,b'\}$. 
	Note that $\pi(w)$ does not contain any  $a$'s or $b$'s. 
	Clearly, 
	$(\loc^\A,\vect{d}),w \sim_\pi(\loc^\A,\vect{d}),\pi(w)$. By Corollary \ref{corollary:gnra_bad}, there still exists an accepting run of $\A$ on $\pi(w)$ that starts in $(\loc^\A,\vect{d})$. 	
	We prove that $\succ_\B(C,\pi(w))$ is non-accepting. 
	Let $(\loc,c)\in C$. 
	We distinguish three cases. 
	\begin{enumerate}
	\item Let $c\not\in\{b,b'\}$. 
	Then $(\loc,c),w\sim_\pi (\loc,c),\pi(w)$. 
	Since $\succ_\B((\loc,c),w)$ is non-accepting by assumption, so that by Corollary \ref{corollary:gnra_bad} also $\succ_\B((\loc,c),\pi(w))$ is non-accepting. 
	
	\item Let $c=b$. By $\indiscernible{a}{b}{\config}$, the state $(\loc,a)$ is in $C$ and $(\loc, a),\pi(w)\sim (\loc,c),\pi(w)$
	since $a$ and $c$ do not appear in $w$. By essential coverability of $\config$, $\{(\loc,a),(\loc,c)\}\subseteq\config$ is coverable. 
	By Corollary \ref{corollary:gura_bad} we obtain that $\succ_\B((\loc,c),\pi(w))$ is non-accepting.

	\item Let $c=b'$. Note that $(\loc,b),w \sim_\pi (\loc,b'),\pi(w)$.  
	Recall that $b'\not\in\supp(\config)$.
	This implies that $\data(\config\cap(\{\loc\}\times\domain_\bot))$ is cofinite. 
	We distinguish two cases. 
	\begin{itemize}
	\item $b\in\data(\config\cap(\{\loc\}\times\domain_\bot))$, i.e., $(\loc,b)\in\config$. But note that $(\loc,b)\not\in\config^+_b$ by cofiniteness of $\data(\config\cap(\{\loc\}\times\domain_\bot))$.
	Hence $(\loc,b)\in \config\backslash\{(\loc^+,b)\}$.  
	\item  $b\not\in\data(\config\cap(\{\loc\}\times\domain_\bot))$, i.e., $(\loc,b)\in\config^-_b$. 
	\end{itemize}
	In both cases, we have proved above that $\succ((\loc,b),w)$ is non-accepting. By  $(\loc,b),w \sim_\pi (\loc,b'),\pi(w)$ and Corollary \ref{corollary:gnra_bad}, $\succ_\B((\loc,b'),\pi(w))$ is non-accepting, too.
	\end{enumerate}
	Altogether we have proved that $\succ_\B(\config,\pi(w))$ is non-accepting, while there exists some accepting run of $\A$ on $\pi(w)$  starting in $(\loc^\A,\vect{d})$. 
	This concludes the proof for the $(\Leftarrow)$-direction.

	\medskip
	
	$(\Rightarrow)$ 
	Suppose there exists some data word $w$ such that  there exists some accepting run of $\A$ on $w$ starting in $(\loc^\A,\vect{d})$, and $\succ_\B(\config,w)$ is non-accepting. 
	We assume in the following that $\succ_\B(\config\setminus \config_b^+\cup \config_b^-,w)$ is accepting; otherwise we are done. 
	Let $(\loc^-,b)$ be a state in $\config_b^-$ such that $\succ_\B((\loc^-,b),w)$ is accepting. 
	Pick some datum $a'\in\domain_\bot$ such that $a'\not\in\data(w) \cup \supp(\config)\cup\data(\vect{d})$. 
	Let $\pi$ be the isomorphism defined by $\pi(b)=a$, $\pi(a)=a'$, $\pi(a')=b$, and $\pi(d)=d$ for all $d\in\domain\backslash\{a,b,a'\}$. 
	Clearly, $(\loc^\A,\vect{d}),w \sim_\pi(\loc^\A,\vect{d}),\pi(w)$, so that by Corollary \ref{corollary:gnra_bad},  there exists some accepting run of $\A$ on $\pi(w)$ starting in $(\loc^\A,\vect{d})$. 
	We prove that $\succ_\B(\config\backslash\config^+_b\cup\config^-_b,\pi(w))$ is non-accepting. 
	Let $(\loc,c)\in \config\backslash\config^+_b\cup\config^-_b$. We distinguish the following cases:
	\begin{enumerate}
	\item Let $c=a$, i.e., $(\loc,a)\in\config$. 
	By $\indiscernible{a}{b}{\sconfig}$, we also have $(\loc,b)\in\config$. 
	Note that $(\loc,b),w\sim_\pi(\loc,a),\pi(w)$. 
	Note that $(\loc,b)\neq (\loc^-,b)$ by Fact \ref{fact:minusandconfignointersect}. 
	By assumption, $\succ_\B((\loc,b),w)$ is non-accepting. 
	By Corollary \ref{corollary:gura_bad},  $\succ_\B((\loc,a),\pi(w))$ is non-accepting, too. 
	\item Let $c\neq a$. 
	Note that also $(\loc^-,b),w \sim_\pi (\loc^-,a),\pi(w)$. 
	Recall that $\succ_\B((\loc^-,b),w)$ is accepting. 
	By Corollary \ref{corollary:gnra_bad},  $\succ_\B((\loc^-,a),\pi(w))$ is accepting. 
	We prove below that $\{(\loc^-,a),(\loc,c)\}$ is coverable. Proposition \ref{prop:gura_implied_badness} then implies that
	 $\succ_\B((\loc,c),\pi(w))$ is non-accepting.

	 Recall that $\data((\{\loc^-\}\times\domain)\cap C)$ is cofinite. 
	 Pick some datum $d\in\domain\backslash\{c\}$ such that $(\loc^-,d)\in\config$. We distinguish two cases. 
	 
	 \begin{itemize}
	 \item Assume 
	 $(\loc,c)\in\config\backslash\config^+_b$. 
	 Since $\config$ is essentially coverable, 
	 the set $\{(\loc^-,d), (\loc,c)\}$ is coverable. 
	 Hence there must exist some data word $u$ such that $\{(\loc^-,d),(\loc,c)\}\subseteq \succ_\B((\loc_\init,\bot),u)$. 
	Let $\pi'$ be a partial isomorphism satisfying $\pi'(d)=a$, $\pi'(a)=d$,  and $\pi'(e)=e$ for all $e\in\data(u)\cup\{c\}$. 
	Then $\{(\loc^-,a),(\loc,c)\}\subseteq \succ_\B((\loc_\init,\bot),\pi'(u))$, hence $\{(\loc^-,a),(\loc,c)\}$ is coverable. 

	\item  Second suppose $(\loc,c)\in\config^-_b$, i.e., $c=b$. 
	 This implies that $\data(\config \cap (\{\loc\}\times\domain))$ is cofinite. 
	 Pick some datum $e\in\domain\backslash\{d\}$ such that $(\loc,e)\in\config$. 
	 Since $\config$ is essentially coverable, 
	 the set $\{(\loc^-,d),(\loc,e)\}$ is coverable. 
	 Hence there must exist some data word $u$ such that $\{(\loc^-,d),(\loc,e)\}\subseteq\succ_\B((\loc_\init,\bot),u)$. 
	 Let $\pi'$ be a partial isomorphism satisfying $\pi'(d)=a$, $\pi'(a)=d$, $\pi'(b)=e$, $\pi'(e)=b$, and $\pi'(f)=f$ for all $f\in\data(u)$. Then $\{(\loc,b),(\loc^-,a)\}\subseteq \succ_\B((\loc_\init,\bot),\pi'(u))$, hence $\{(\loc,c),(\loc^-,a)\}$ is coverable.
	 \end{itemize}
	\end{enumerate}
	Altogether we have proved that $\succ_\B((\config\backslash\config^+_b)\cup\config^-_b,\pi(w))$ is non-accepting, while there is an accepting run of $\A$ on $\pi(w)$ starting in $(\loc^\A,\vect{d})$. 
	This finishes the proof for the $(\Rightarrow)$-direction, and thus the proof of the Proposition. 	
\end{proof}

\subsection{The Algorithm}
When a synchronized configuration $\sconfig'$ is obtained from some essentially coverable synchronized configuration $\sconfig=((\loc,\vect{d}),\config)$ by applying Proposition~\ref{prop:collapse_gura} to two distinct data values $a,b\in\supp(\config)$, we say that $\sconfig$ \emph{collapses to} $\sconfig'$.
We say that $\sconfig$ is \emph{maximally collapsed} if one cannot find two distinct data values $a,b\in\supp(\config)$ that satisfy the assumptions of Proposition~\ref{prop:collapse_gura}.
Note that, by Proposition \ref{prop:esscov},  the synchronized configuration $\sconfig'$ in Proposition~\ref{prop:collapse_gura} is again essentially coverable.
By iterating Proposition~\ref{prop:collapse_gura}, one obtains that an essentially coverable synchronized configuration reaches a bad synchronized configuration
if, and only if, it collapses in finitely many steps to a maximally collapsed synchronized configuration that also reaches a bad synchronized configuration.

The number of maximally collapsed configurations is asymptotically bounded
by $2^{k\log(k)2^{|\locs|}}$. Indeed, a maximally collapsed configuration $((\loc,\vect d),\config)$ can be recovered up to $\sim$ by:
\begin{itemize}
	\item The location $\loc$ and the equivalence class of $\vect d$,
	\item A list $L_\bot,L_1,\dots,L_k$ of subsets of $\locs$,
	\item A set $\{L_{k+1},\dots,L_p\}$ of subsets of $\locs$,
	\item For each location $\loc\in\locs$, a bit $b_\loc\in\{0,1\}$.
\end{itemize}
From this, one can constitute a configuration $\sconfig=((\loc,\vect d'), \config)$ where:
\begin{itemize}
	\item $\vect d'$ is an arbitrary tuple in the equivalence class of $\vect d$, using only data from $\{\bot,1,\dots,k\}$, 
	\item For every $i\in\{\bot,1,\dots,k\}$ and $\loc'\in L_i$, $\config$ contains $(\loc',d'_i)$,
	\item For every $i\in\{k+1,\dots,p\}$ and $\loc'\in L_i$, $\config$ contains $(\loc',i)$,
	\item For each $d\in\domain\setminus\{1,\dots,p\}$, $(\loc',d)$ is in $\config$ iff $b_{\loc'}=1$. That is, the bit $b_{\loc'}$ is set to 1 to indicate that $(\{\loc'\}\times\domain) \cap C$ is cofinite.
\end{itemize}
Thus, one can bound the number of maximally covered configurations by $|\locs|\times k^k \times (k+1)2^{|\locs|} \times 2^{2^{|\locs|}}\times 2^{|\locs|}$
which is asymptotically $2^{k\log(k)2^{|\locs|}}$.
Consider the graph whose vertices are the maximally collapsed synchronized configurations and which contains an edge $\sconfig\leadsto\sconfig'$ iff there exists an $\sconfig''$
such that $\sconfig\sTo\sconfig''$ and $\sconfig''$ collapses to $\sconfig'$.
This graph  has doubly-exponential size in $\A$ and $\B$, and the relation $\leadsto$ can be decided in polynomial space~\cite{DBLP:conf/stacs/MottetQ19}.
Thus, one obtains that the reachability problem in this graph can be decided in exponential space, so that the containment problem for 1-register \GURA\  is in \EXPSPACE.

\begin{theorem}
The containment problem $L(\A)\subseteq L(\B)$ is in $\textup{\EXPSPACE}$, if $\A$ is a $\GRA$ and $\B$ is an unambiguous $\GRA$ with a single register. 
\end{theorem}

\bibliography{IEEEabrv,../lit.bib}

\begin{thebibliography}{1}

\bibitem{DBLP:conf/dcfs/Colcombet15}
T.~Colcombet.
\newblock Unambiguity in automata theory.
\newblock In J.~Shallit and A.~Okhotin, editors, {\em Descriptional Complexity
  of Formal Systems - 17th International Workshop, {DCFS} 2015, Waterloo, ON,
  Canada, June 25-27, 2015. Proceedings}, volume 9118 of {\em Lecture Notes in
  Computer Science}, pages 3--18. Springer, 2015.

\bibitem{DBLP:journals/tocl/DemriL09}
S.~Demri and R.~Lazic.
\newblock {LTL} with the freeze quantifier and register automata.
\newblock {\em {ACM} Trans. Comput. Log.}, 10(3), 2009.

\bibitem{DBLP:journals/tcs/KaminskiF94}
M.~Kaminski and N.~Francez.
\newblock Finite-memory automata.
\newblock {\em Theor. Comput. Sci.}, 134(2):329--363, 1994.

\bibitem{KaminskiZeitlin}
M.~Kaminski and D.~Zeitlin.
\newblock Finite-memory automata with non-deterministic reassignment.
\newblock {\em International Journal of Foundations of Computer Science},
  {Volume 21, Issue 05}, 2010.

\bibitem{DBLP:conf/stacs/MottetQ19}
A.~Mottet and K.~Quaas.
\newblock The containment problem for unambiguous register automata.
\newblock In R.~Niedermeier and C.~Paul, editors, {\em 36th International
  Symposium on Theoretical Aspects of Computer Science, {STACS} 2019, March
  13-16, 2019, Berlin, Germany}, volume 126 of {\em LIPIcs}, pages 53:1--53:15.
  Schloss Dagstuhl - Leibniz-Zentrum fuer Informatik, 2019.

\bibitem{DBLP:journals/tocl/NevenSV04}
F.~Neven, T.~Schwentick, and V.~Vianu.
\newblock Finite state machines for strings over infinite alphabets.
\newblock {\em {ACM} Trans. Comput. Log.}, 5(3):403--435, 2004.

\bibitem{DBLP:conf/lics/OuaknineW04}
J.~Ouaknine and J.~Worrell.
\newblock On the language inclusion problem for timed automata: Closing a
  decidability gap.
\newblock In {\em 19th {IEEE} Symposium on Logic in Computer Science {(LICS}
  2004), 14-17 July 2004, Turku, Finland, Proceedings}, pages 54--63. {IEEE}
  Computer Society, 2004.

\bibitem{DBLP:journals/tocl/QuaasS19}
K.~Quaas and M.~Shirmohammadi.
\newblock Synchronizing data words for register automata.
\newblock {\em {ACM} Trans. Comput. Log.}, 20(2):11:1--11:27, 2019.

\end{thebibliography}

\end{document}